\pdfoutput=1
\documentclass[10pt,journal,compsoc]{IEEEtran}
\usepackage{url}
\usepackage[pdftex]{graphicx}
\usepackage{amsmath}
\usepackage{biblatex}
\bibliography{bib.bib}
\begin{document}

\title{DisTRaC: Accelerating High Performance Compute Processing for Temporary Data Storage}
\author{Gabryel~Mason-Williams,
        Dave~Bond,
        and~Mark~Basham
\IEEEcompsocitemizethanks{\IEEEcompsocthanksitem G. Mason-Williams was with Diamond Light Source, Scientific Computing, SSCC, Didcot, Oxfordshire, UK and studies at the University of Plymouth, School of Engineering, Computing and Mathematics, Plymouth, Devon, UK. E-mail: gabryel.research@gmail.com.\protect
\IEEEcompsocthanksitem D. Bond is with Diamond Light Source, Scientific Computing, SSCC, Didcot, Oxfordshire, UK. E-mail: dave.bond@diamond.ac.uk.\protect
\IEEEcompsocthanksitem M. Basham is with the Rosalind Franklin Institute, Artificial Intelligence and Informatics, Didcot, Oxfordshire, UK and Diamond Light Source, Data Analysis Group, SSCC, Didcot, Oxfordshire, UK. E-mail: mark.basham@rfi.ac.uk.\protect\\}
}
\markboth{}%
{}
\IEEEtitleabstractindextext{%
\begin{abstract}

High Performance Compute (HPC) clusters often produce intermediate files as part of code execution and message passing is not always possible to supply data to these cluster jobs. In these cases, I/O goes back to central distributed storage to allow cross node data sharing. These systems are often high performance and characterised by their high cost per TB and sensitivity to workload type such as being tuned to small or large file I/O. However, compute nodes often have large amounts of RAM, so when dealing with intermediate files where longevity or reliability of the system is not as important, local RAM disks can be used to obtain performance benefits. In this paper we show how this problem was tackled by creating a RAM block that could interact with the object storage system Ceph, as well as creating a deployment tool to deploy Ceph on HPC infrastructure effectively. This work resulted in a system that was more performant than the central high performance distributed storage system used at Diamond reducing I/O overhead and processing time for Savu, a tomography data processing application, by 81.04\% and 8.32\% respectively.
\end{abstract}

}

\maketitle

\section{Introduction}
\label{sec:introduction}
\IEEEPARstart{D}{iamond} 
Light Source (Diamond) is the UK's national synchrotron facility, it collects and processes an immense amount of data a year; High Performance Compute (HPC) clusters process this data. HPC clusters often produce intermediate files as part of code execution, message passing is not always possible to supply data to these cluster jobs especially when doing High Throughput Computing (HTC). In these cases, I/O goes back to central high performance distributed storage to allow cross node data sharing. These systems often are characterised by their high cost per TB and sensitivity to workload type such as being tuned to small or large file I/O. However, compute nodes often have large amounts of RAM, so when dealing with intermediate files where longevity or reliability of the system is not as important local RAM disks are often used for the performance benefits. This paper explores the possibility of combining these two ideas, to provide a fast RAM disk, which is also available as a shared resource through the exploitation of object storage. Use of technologies such as these reduce the I/O cost of a data pipeline and can reduce processing times while requiring no additional infrastructure. 

\section{Object Storage}

Object storage is a way to store and handle unstructured data,  such as images, audio files, videos and data that that does not have a predefined way of storing it sensibly. The data is stored as an object with three parts, the data, metadata (that describes information about it) and a unique identifier. 

Metadata allows for the objects to be stored in a flat structure called a pool, unlike the conventional hierarchy storage structure of file storage where directories and subdirectories are used. Using a storage pool can make data retrieval quicker. The flat structure also allows for near infinite horizontal scaling by adding storage devices and nodes to the system. This in turn allows for more processing and support of higher throughput rates as required.  Object storage can cope with a variety of object sizes with less tuning than conventional file storage as object storage provides chunking functionality. Chunking allows large objects to be split into smaller objects which can improve the performance of the system.

The communication protocol for object storage is via a REST API as each object has a unique identifier, and there are no folders or files to transverse. The use of a REST API makes object storage cloud-native, with common object storage interaction supported via the S3 \cite{s3} or Swift \cite{swift} API's.

\section{RAM Ceph}

Ceph is a free and open source storage system that can act as a file, object and block storage \cite{ceph}.  To implement Ceph as an object storage cluster, it requires monitor nodes (MON)s,  manager nodes (MGR)s (since the Luminous version 12.*), Ceph Object Storage Devices (ODS)s and Rados Gateways (RGW)s (if required). MONs are used to monitor the state of the cluster and maintains the master copy for the cluster map, MGRs provide extra monitoring as well as interfaces to external systems for additional monitoring, OSDs interact with a logical disk to store data and RGWs interact with the cluster via S3 or Swift.

Ceph requires block devices as the storage medium such as Hard Disk Drives (HDDs) or Non-Volatile Memory Express (NVMe). The need for block devices derives from Ceph's use of the Logical Volume Management (LVM) system to create physical and logical volumes for OSDs. Ceph is unable to use RAM in its default state as it presented as system memory, not as a block device. Changing RAM to a block device requires a change to the way RAM is present within the system.

Linux provides two kernel modules that convert the representation of RAM from system memory to a block device, BRD and ZRAM \cite{zram}.  The difference between BRD and ZRAM  is that ZRAM creates a compressed RAM device that defaults to LZO compression.  ZRAM proved more performant than BRD, when tested using dd and averaged over three runs, as shown in Table \ref{tab:read-zramvsgramvsbrd} and \ref{tab:write-zramvsgravsbrd}. 

ZRAM's use of LZO would inevitably affect the performance of using a RAM object store to process intermediate data as it will require CPU resources to compress and decompress data, rather than processing constantly. To resolve the issue of compression and to maintain performance General RAM (GRAM) was created \cite{GRAM}. GRAM is a kernel module based on ZRAM; omitting ZRAM's compression aspect, but maintaining the same read and write mechanism, allows GRAM to no longer require the excess CPU time to compress and decompress data. Consequently, GRAM is more performant than BRD, when tested using dd and averaged over three runs, as shown in Table \ref{tab:read-zramvsgramvsbrd} and \ref{tab:write-zramvsgravsbrd}.  

\begin{table}[ht]
\caption{Read throughput of ZRAM, GRAM and BRD in GB/s}
\label{tab:read-zramvsgramvsbrd}
\begin{tabular}{llll}
Block Size & ZRAM              & GRAM              & BRD               \\
\hline
4K         & $\pmb{1.623 \pm 0.005}$ & $1.614 \pm 0.002$ & $1.483 \pm 0.109$ \\
40K        & $\pmb{2.194 \pm 0.006}$ & $2.188 \pm 0.004$ & $1.823 \pm 0.017$ \\
400K       & $\pmb{2.193 \pm 0.023}$ & $2.154 \pm 0.020$ & $1.824 \pm 0.004$ \\
4M         & $2.124 \pm 0.007$ & $\pmb{2.130 \pm 0.012}$ & $1.756 \pm 0.009$ \\
40M        & $\pmb{1.737 \pm 0.007}$ & $1.736 \pm 0.005$ & $1.493 \pm 0.012$ \\
400M       & $\pmb{1.715 \pm 0.011}$ & $\pmb{1.715 \pm 0.011}$ & $1.494 \pm 0.013$  
\end{tabular}
\end{table}

\begin{table}[ht]
\caption{Write throughput of ZRAM, GRAM and BRD in GB/s}
\label{tab:write-zramvsgravsbrd}
\begin{tabular}{llll}
Block Size & ZRAM              & GRAM               & BRD               \\
\hline
4K         & $\pmb{1.175 \pm 0.007}$ & $1.144 \pm 0.023$  & $1.090 \pm 0.050$ \\
40K        & $\pmb{1.389 \pm 0.007}$ & $1.384 \pm 0.018$  & $1.302 \pm 0.010$ \\
400K       & $\pmb{1.386 \pm 0.041}$ & $1.381 \pm 0.021$  & $1.327 \pm 0.065$ \\
4M         & $1.409 \pm 0.009$ & $\pmb{1.420 \pm 0.017}$  & $1.261 \pm 0.116$ \\
40M        & $\pmb{1.310 \pm 0.011}$ & $1.280 \pm 0.039$  & $1.197 \pm 0.041$ \\
400M       & $\pmb{1.301 \pm 0.026}$ & $1.276 \pm 0.030$  & $1.205 \pm 0.010$  
\end{tabular}
\end{table}

GRAM is not a registered block device type in the Linux kernel. As a result, to create physical and logical volumes, GRAM needs modification of the LVM configuration file. The addition of GRAM to the allowed types in the LVM configuration file results in Ceph supporting GRAM (which omits compression entirely), allowing for a performant RAM-based OSD.

\section{Scalable deployment}

Ceph has numerous deployment tools that work sequentially; requiring SSH for the deployment. The use of SSH negatively impacts the stability of the HPC cluster when deploying on to HPC infrastructure that uses a resource manager such as Univa Grid Engine (UGE) for job allocation \cite{uge}. Sequential use of SSH is an inefficient way to scale in a short amount of time as it increases deployment time which increases the amount of time to process the data, therefore, negatively affects using Ceph as a distributed transient RAM-based object-store. Production-based Ceph deployment tools add a delay when deploying MONs, allowing ample time for the system to find quorum, which increases overall deployment time, highlighting its inefficiencies when applied to a distributed transient RAM-based object-store. Another inefficiency highlighted when applying Ceph deployment tools to create a distributed transient RAM based object store, was that it sets storage pool replication to three for data redundancy, reducing the overall performance of the system as well as requiring at least three OSDs. To remove these inefficiencies and issues, DisTRaC, a bespoke distributed transient RAM Ceph deployment tool, was created \cite{distrac}. DisTRaC takes full advantage of the HPC environment, where a Network File System (NFS) is present to create a RAM-based Ceph instance.

UGE allocates resources (slots) to jobs running within a cluster. When applications SSH to a host within a UGE controlled cluster, the application is out of UGE's scope, this means that an application is likely to take resources that UGE has allocated to a running job. Stealing resource has an adverse effect on parallel or single host jobs running because it can interrupt the job running. To run jobs simultaneously across multiple nodes, UGE requires a parallel environment (PE). Creating a PE to allocate one slot per host for DisTRaC means that when DisTraC is in the HPC environment, it uses only the resources allocated to the job; removing the issue of SSHing and potentially stealing another job/s resources as UGE allocates the proportional resources required for DisTRaC.

\begin{figure}[htb!]
\centering
\includegraphics[width=\columnwidth]{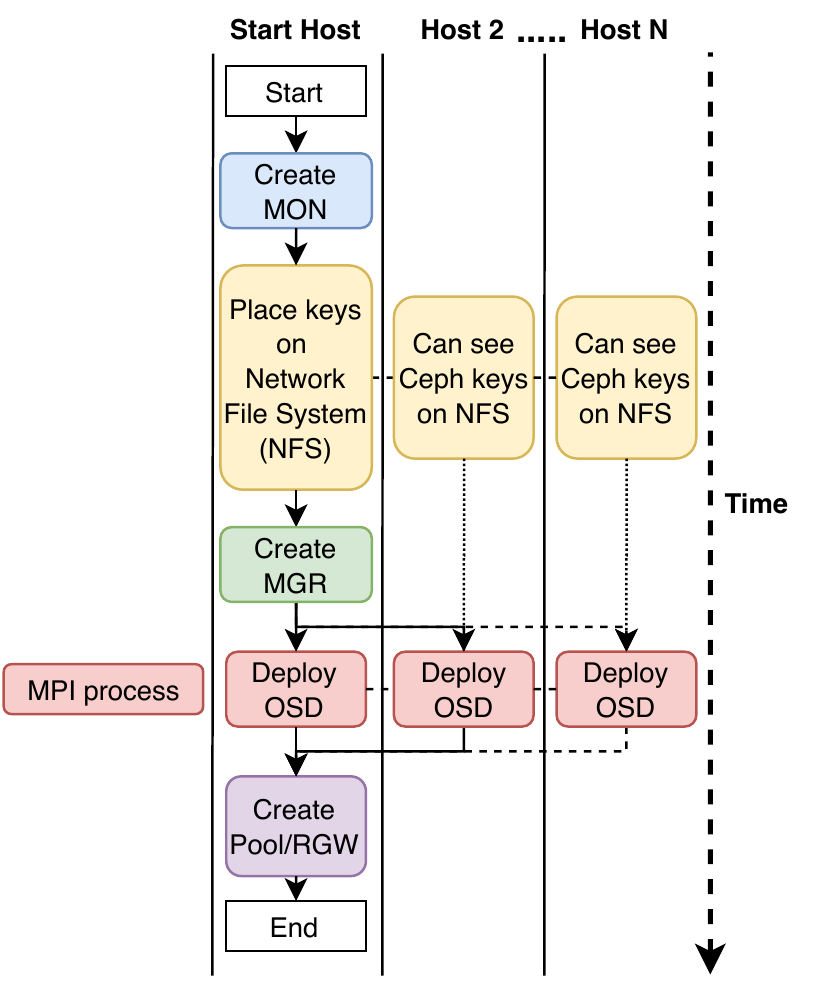}
\caption{DisTRaC deployment - A MON is created on the head node first; then the keys are put on the NFS. Putting the keys on NFS means the other hosts can see the keys. Then a MGR is created as required by Ceph Luminous version 12.*. MPI is then run across the specified hosts using the keys on NFS to create GRAM and the subsequent OSDs in parallel. The final step either creates a pool or a RGW, depending on what the user specified.}
\label{fig:distrac-deploy}
\end{figure}

\begin{figure}[htb!]
\centering
\includegraphics[width=\columnwidth]{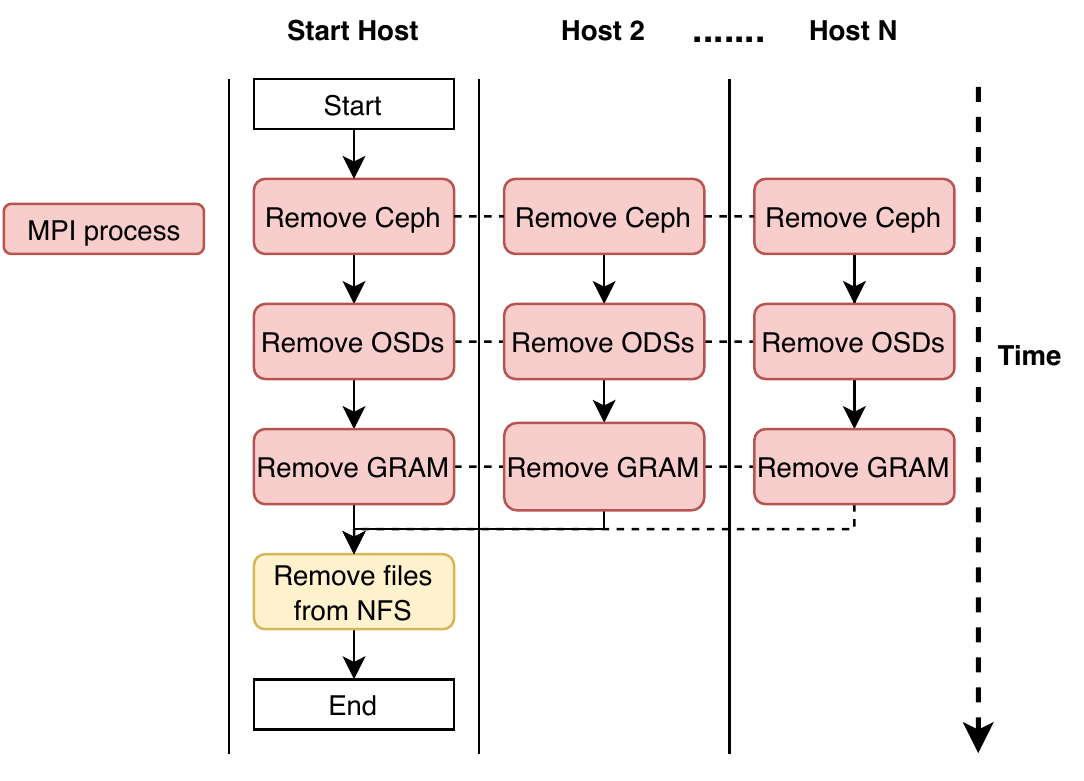}
\caption{DisTRaC removal - MPI is run across the hosts to remove Ceph's daemons and clears the Ceph system folders. Then the MPI is run to remove the OSDs, which removes the ceph volume groups, unmounts the OSDs from the system and clears the OSDs system files. Then MPI is run to remove GRAM from the system freeing up the RAM used in the process. Finally, the keys are removed from NFS.}
\label{fig:distrac-remove}
\end{figure}

When a Ceph instance is being created a set of keys for MONs, MGRs, OSDs, Clients and RGWs (if required) are made. In a traditional deployment, the system would SSH to the hosts copying the keys to start the relevant daemon and check the state of the system. This step is unrequired in an HPC environment where NFS is present, as hosts can already see the keys once made. Instead, the same action of starting the relevant daemon is needed. DisTRac takes full advantage of homogeneous nature of HPC systems and the PE to run the jobs in parallel using Message Passing Interface MPI \cite{walker1996mpi}. Taking full advantage of this means DisTRac efficiently deploys OSDs on to the appropriate nodes in parallel, bypassing the issue of sequential SSH deployment as well as the need of passwordless SSH. Afterwards, a RGW on the host node or storage pool is created, see Fig. \ref{fig:distrac-deploy} for example deployment. 
The removal of Ceph using DisTRaC follows a similar vein as deployment, using MPI to remove Ceph daemons and processes, then removing the keys and files from NFS see Fig. \ref{fig:distrac-remove}.

DisTRaC builds a RAM-based Ceph instance. The Ceph instance does not need to find quorum with multiple MONs because the system is inherently volatile due to using RAM as the storage device. Deploying a singular MON removes the need to check if the system reaches quorum, reducing the deployment time.  Ceph stores data with redundancy by default; as a result, data replication is set to three. A data replication setting of three means data is copied and stored across three different OSDs. This redundancy impacts the performance of the system as it is dealing with three times the amount of data and requires at least three OSDs to handle the data as the replicated data is situated on different OSDs to the on containing the original data. RAM-based Ceph is focused on performance and has limited storage capacity, meaning DisTRaC sets the replication of the system to one, creating a more performant system that only stores the data required. Meaning there is zero fault tolerance. However, this is not an issue as the data stored is intermediate data.

Due to the effective use of UGE, PE and MPI DisTRaC deploys and removes a RAM-based Ceph cluster within 120 seconds irrespective of cluster size. See Table \ref{tab:deployment-time} when deploying one OSD per node and only creating one pool. A short deployment time made this tool a viable solution for creating a distributed transient RAM based Ceph cluster instance as it will have little impact on the overall processing time of applications using it. 

\begin{table}[htb]
\caption{Deployment time in seconds of one OSD per node with a storage pool}
\label{tab:deployment-time}
\begin{tabular}{llll}
NODES & Deploy           & Remove           & Total             \\
\hline
1     & $20.766\pm0.059$ & $93.723\pm0.133$ & $114.488\pm0.146$ \\
2     & $22.011\pm1.384$ & $95.754\pm2.432$ & $117.765\pm3.169$ \\
3     & $22.779\pm0.885$ & $96.598\pm2.244$ & $119.377\pm2.482$ \\
4     & $23.732\pm1.515$ & $95.620\pm0.396$ & $119.353\pm1.425$ \\
5     & $22.404\pm0.764$ & $96.514\pm0.651$ & $118.918\pm0.515$ \\
6     & $22.946\pm0.680$ & $95.534\pm0.542$ & $118.480\pm1.022$
\end{tabular}%
\end{table}

\section{Savu use case}

\begin{figure}[ht]
\centering
\includegraphics[width=\columnwidth]{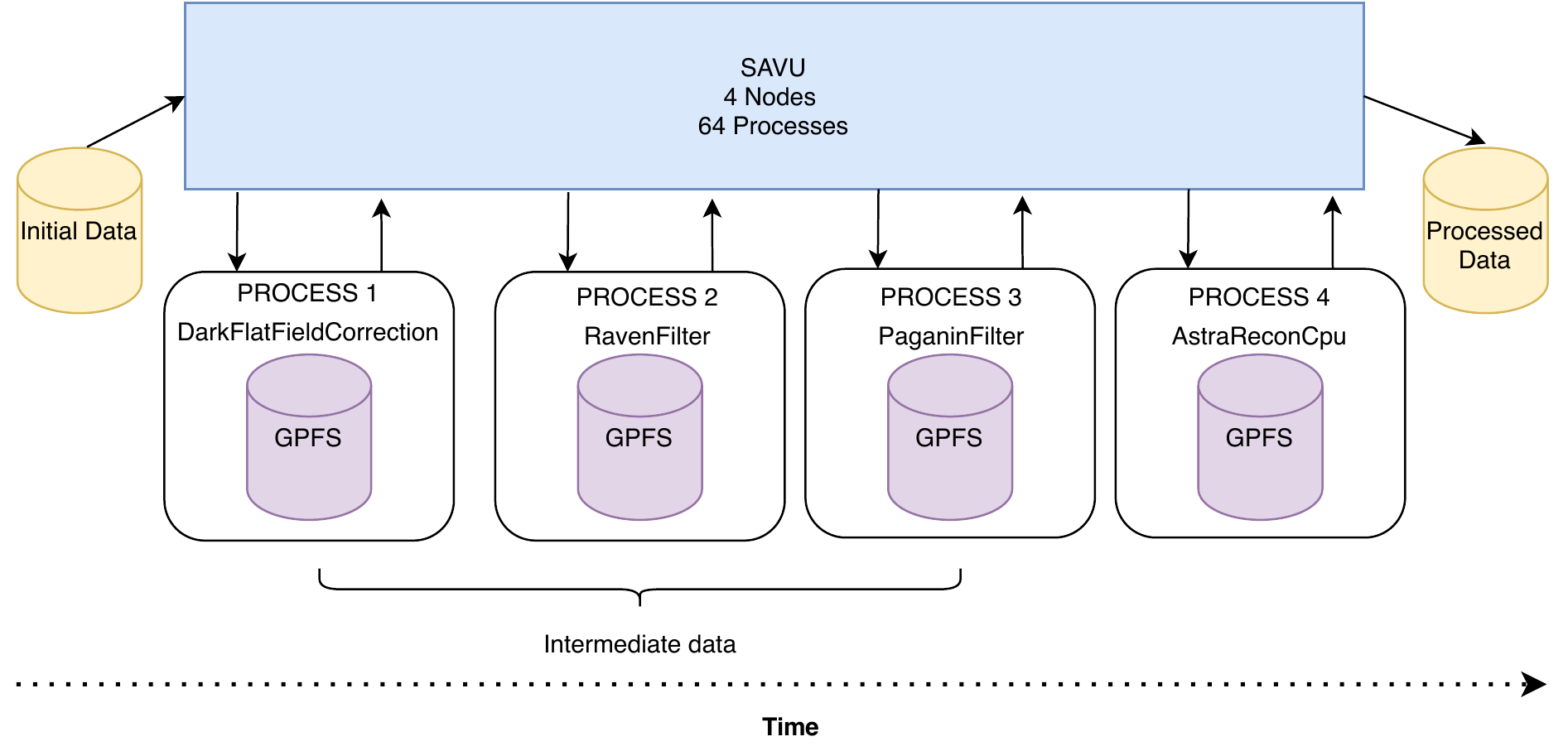}
\caption{Savu data pipeline - Initial data is read and processed on the compute nodes for process one. During this, the processed data is written to GPFS. Afterwards, that data is read from GPFS to the compute nodes for process two with the processed data written to GPFS as well. The same process occurs for process three and four.  The last step then stores the processed data from all steps.}
\label{fig:savu}
\end{figure}

Savu is a Python application that uses MPI to effectively process huge tomographic data-sets for synchrotron beamlines \cite{wadeson2016savu, savu24}.  A typical data process consists of several image correction and filtering steps prior to the final reconstruction step. These processing steps produce a significant amount of data and Savu traditionally uses  MPIIO and the HDF5 \cite{hdf5} libraries to store the intermediate data created during processing as well as the final reconstruction data on the Diamond central high-performance filesystem GFPS \cite{gpfs} as shown in Fig. \ref{fig:savu}.

During the EU SAGE project \cite{Sage} the DosNa \cite{DosNa} library was created to allow Savu to interface with object stores, as object stores showed promise for some areas of the processing chain. Albeit, Savu currently uses a central high performance distributed storage system to store large data, the total overhead required is only several hundred gigabytes, which is well within the memory availability of the cluster nodes which run the task at Diamond. 

Here we investigate a Savu job with GPFS across four nodes, 16 slots on each node, a 42.346GB tomographic data-set and a process list of Dark Flat Field Correction, Raven Filter, Paganin Filter and Astra Reconstruction \cite{savu-dataset}. This Savu job takes on average 173.775 minutes process and reconstruct the data, see Table \ref{tab:processing-table}. This job creates a total I/O overhead of 300.900GB with an intermediate data I/O overhead of 243.858GB.

\begin{figure}[ht]
\centering
\includegraphics[width=\columnwidth]{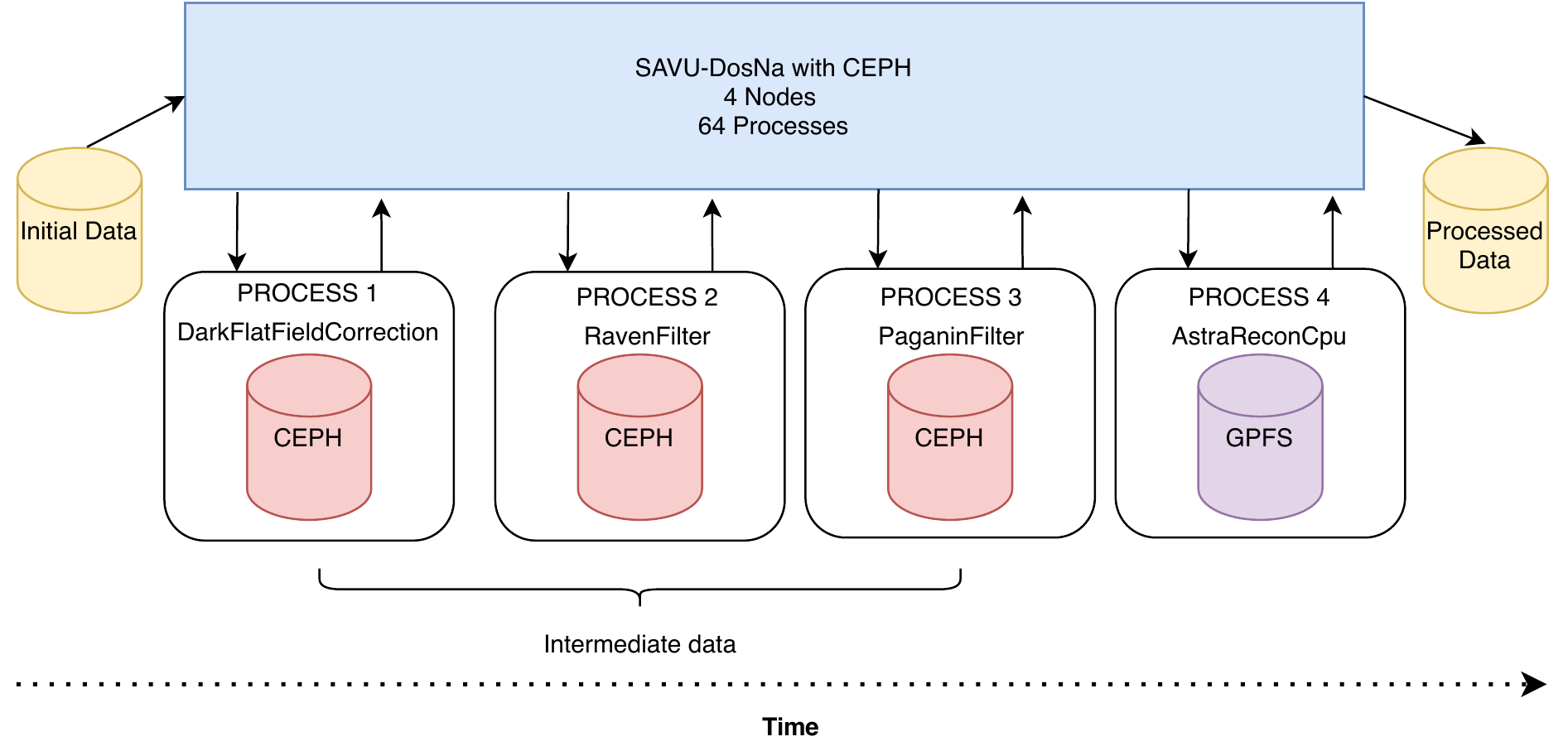}
\caption{Savu-DosNa with Ceph data pipeline - Initial data is read and processed on the compute nodes for process one. During this, the processed data is written to Ceph.  Afterwards, that data is read from Ceph to the compute nodes for process two with the processed data written to Ceph as well. The same occurs for process three. For process four, data is read from Ceph to the compute nodes. However, the processed data is written to GPFS. The last step then only stores the data produced from process four.}
\label{fig:DosNa}
\end{figure}

The total intermediate data overhead is under 400GB, which means DisTRaC and DosNa can be used to process this data within four nodes. The DosNa library, when used with Savu stores all bar the last process in an object-store, AstraReconCpu in this example,  as the last step is not intermediate data. Subsequently, only DarkFlatFieldCorrection, RavenFilter and PaganinFilter will be processed on RAM Ceph as demonstrated in Fig. \ref{fig:DosNa}.

Using the same data-set across four nodes with 16 slots on each node and deploying Ceph using DisTRaC, Savu takes an average total time of 159.324 minutes to process and reconstruct the data, see Table \ref{tab:processing-table}.

\begin{table}[ht]
\caption{Savu and Savu-DosNa with DisTRaC processing time in minutes}
\label{tab:processing-table}
\resizebox{\columnwidth}{!}{%
\begin{tabular}{lll}
Process                 & SAVU              & Savu-DosNa with DisTRaC       \\
\hline
Deploy Ceph             & $0\pm0$           & $0.381\pm0.001$    \\
DarkFlatFieldCorrection & $10.299\pm0.590$  & $2.547\pm0.023$    \\
RavenFilter             & $16.357\pm0.326$  & $2.423\pm0.018$    \\
PaganinFilter           & $13.393\pm0.068$  & $2.501\pm0.025$    \\
AstraReconCpu           & $133.514\pm2.523$ & $149.398\pm2.814$  \\
Remove Ceph             & $0\pm0$           & $1.702\pm0.008$    \\
\hline
Total Job Time          & $173.775\pm2.701$ & $159.324\pm2.790$
\end{tabular}
}
\end{table}

Comparing traditional Savu and Savu-DosNa with DisTRaC, there is a reduction in the overall processing time and I/O overhead of 8.32\% and 81.04\% respectively. Comparatively, process for process there is a reduction of 75.27\%, 85.19\%, 81/32\% for the Dark Flat Field Correction, Raven Filter and Paganin Filter respectively, Astra Recon Cpu has an increased processing time of 11.90\% this is because of the change over from processing with Ceph to GPFS.  This demonstrates that using RAM disks as a shared resource through the medium of object storage is a viable and effective way to reduce I/O overhead as well as decrease processing time.

\section{Extensions of DisTRaC to a more traditional shared temporary file system}

When dealing with sequentially related batch jobs on a cluster, often a shared directory is needed.  As Ceph supports filesystems, future versions of DisTRaC will add this support to allow for a broader range of temporary data processing. Preliminary work has shown this could work in the same way as deploying Ceph as a native object store. 

\section{Future work}

It will be important that future research investigates the potential performance benefits of modifying the chuck size of data being processed when using an object-store. As using a more appropriate chunk size could lead to a further reduction in processing time. It will be worthwhile investigating the use of containers in the deployment of Ceph within DisTRaC as this may reduce the time to deploy and remove Ceph. Future work within the community should focus on adding the functionality to interact with object storage to existing and new applications that would benefit from using the object storage system presented.

\section{Conclusion}

In conclusion, this paper set out to explore the use of RAM disks as a shared resource through the medium of object storage in the hope to reduce the I/O cost of using a central high performance distributed storage system and garner the performance benefits of using RAM. This problem was tackled by creating a RAM block that could interact with the object storage system Ceph, as well as creating a deployment tool to deploy Ceph on HPC infrastructure effectively. This work resulted in a system that was more performant than the central high performance distributed storage system used at Diamond reducing processing time by 8.32\% and I/O overhead by 81.04\%.

Thus, the system presented in this paper can use used to process and handle any form of intermediate data processing that can be stored in an object store. Processing intermediate and temporary data in this way presents a new and innovative way to reduce operational I/O costs and speed up data processing pipelines while utilising existing infrastructure. 

\section*{Acknowledgments}

This work was carried out with the support of the Diamond Light Source, instrument I12 (proposal NT23252) and the authors would like to thank the Ceph user group at Rutherford Appleton Laboratory as well as the Ceph community for their help and advice. 

\printbibliography

\end{document}